\documentclass[prl,twocolumn,
			showkeys,showpacs,
				superscriptaddress]{revtex4}

\usepackage{graphicx}
\usepackage{url}
\usepackage{amsmath}
\usepackage{amssymb}
\usepackage{bm}
\usepackage{dsfont}
\usepackage{graphicx}
\usepackage{subfigure}
\usepackage{dcolumn}

\newcommand{\re}{{\mathrm e}}
\newcommand{\rd}{\mathrm{d}}

\newcommand{\no}{\hat{n}}

\newcommand{\la}{\langle}
\newcommand{\ra}{\rangle}
\newcommand{\tr}{\mathrm{tr}}
\newcommand{\bn}{{\bm n}}

\begin{document}

\title{Generalized Bose-Einstein condensation into multiple states\\ in driven-dissipative systems}

\date{November 20, 2013}

\author{Daniel Vorberg}
\affiliation{Max-Planck-Institut f\"ur Physik komplexer Systeme, N\"othnitzer Str.\ 38, 01187 Dresden, Germany}
\affiliation{Technische Universit\"at Dresden, Institut f\"ur Theoretische Physik, 01062 Dresden, Germany}

\author{Waltraut Wustmann}
\affiliation{Max-Planck-Institut f\"ur Physik komplexer Systeme, N\"othnitzer Str.\ 38, 01187 Dresden, Germany}
\affiliation{Technische Universit\"at Dresden, Institut f\"ur Theoretische Physik, 01062 Dresden, Germany}

\author{Roland Ketzmerick}
\affiliation{Max-Planck-Institut f\"ur Physik komplexer Systeme, N\"othnitzer Str.\ 38, 01187 Dresden, Germany}
\affiliation{Technische Universit\"at Dresden, Institut f\"ur Theoretische Physik, 01062 Dresden, Germany}

\author{Andr\'e Eckardt}
\email{eckardt@pks.mpg.de}
\affiliation{Max-Planck-Institut f\"ur Physik komplexer Systeme, N\"othnitzer Str.\ 38, 01187 Dresden, Germany}

\begin{abstract}
Bose-Einstein condensation, the macroscopic occupation of a single quantum state, appears in equilibrium quantum statistical mechanics and persists also in the hydrodynamic regime close to equilibrium. Here we show that even when a degenerate Bose gas is driven into a steady state far from equilibrium, where the notion of a single-particle ground state becomes meaningless, Bose-Einstein condensation survives in a generalized form: the unambiguous selection of an odd number of states acquiring large occupations. Within mean-field theory we derive a criterion for when a single and when multiple states are \emph{Bose selected} in a non-interacting gas. We study the effect in several driven-dissipative model systems, and propose a quantum switch for heat conductivity based on shifting between one and three selected states.
\end{abstract}

\pacs{05.30.Jp, 67.10.Ba, 67.85.Jk}
\keywords{Bose-Einstein condensation, ideal Bose gas, driven-dissipative quantum systems,
non-equilibrium steady state, quantum Markov process, Floquet systems}







\maketitle

In quantum many-body physics there is currently a huge interest in non-equilibrium phenomena
beyond the hydrodynamic description of systems retaining approximate local equilibrium.
Recently, intriguing results have been obtained for paradigmatic scenarios: the dynamics away
from equilibrium in response to a parameter variation 
\cite{CampisiEtAl11,Dziarmaga10,PolkovnikovEtAl11}, the possible relaxation towards
equilibrium \cite{Dziarmaga10,PolkovnikovEtAl11} versus many-body localization
\cite{BaskoEtAl06,HuseEtAl13}, and the control of many-body physics by means of strong periodic
forcing \cite{EckardtEtAl05,ZenesiniEtAl09, StruckEtAl11}. Another fundamental scenario of
quantum many-body dynamics are non-equilibrium steady states of driven-dissipative systems,
with transport of, e.g., mass or energy through the system \cite{PeriEtAl03,ZiaSchmittmann07,DharEtAl12}. 
In this context one might ask simple questions:
What are the properties of an ideal Bose gas driven to a steady state far from equilibrium? 
In particular, what happens in the quantum degenerate regime, where in equilibrium Bose-Einstein
condensation occurs?

In this letter we investigate the quantum degenerate regime of driven-dissipative ideal Bose gases
of $N$ particles in steady states far from equilibrium, assuming weak coupling to the environment.
Examples of such systems comprise bosons coupled to two heat baths of different temperature and
time-periodically forced bosons in contact with a single heat bath. For large densities these
systems are found to exhibit an intriguing generic behavior, namely the single-particle states
unambiguously separate into two groups: one, that we call
\emph{Bose selected}, whose occupations increase linearly when the total particle number is 
increased at fixed system size, and another one whose occupations saturate.
Remarkably, this generalized form of Bose condensation is a very consequence of bosonic 
indistinguishability, not relying on thermodynamic equilibrium. 
We show examples both with the number of selected states being extensive and of order one; the latter
corresponding to a fragmented condensate \cite{MuellerEtAl06} with a macroscopic occupation of each
selected state (not relying on ground-state degeneracy). 
We propose to switch the heat conductivity of a system by shifting between one selected state
(corresponding to standard Bose condensation) and three selected states. 

Our findings are relevant for artificial many-body quantum systems such as superconducting
and optical circuits \cite{HartmannEtAl08,SchoelkopfGirvin08,HoukEtAl12,
CarusottoCiuti13}, exciton-polariton fluids \cite{DentEtAl10,CarusottoCiuti13}, or photons in 
a dye-filled cavity \cite{KlaersEtAl10}, that are intrinsically driven-dissipative. 
Tailored dissipation has also been used or 
proposed as a powerful tool for quantum engineering in ultracold atomic quantum gases
\cite{SyassenEtAl08,RoncagliaEtAl10,MuellerEtAl12,BarontiniEtAl13} and trapped ions
\cite{MuellerEtAl12,BermudezEtAl13,SchindlerEtAl13}. 
Our results, moreover, provide a connection between Bose condensation in quantum systems
and the phenomenon of real-space condensation in classical non-equilibrium models
\cite{BianconiBarabasi01,EvansHanney05,HirschbergEtAl09,GrosskinskyEtAl11,Schadschneider11}, where also
condensation into multiple states has been found
\cite{EvansEtAl06,ScharzkopfEtAl08,KimEtAl10,ThompsonEtAl10}.

Consider an open quantum system of a single particle weakly coupled to
an environment, with reduced density operator $\rho$.
The dynamical map for the time evolution of $\rho$ shall be given by a Markovian 
master equation $\dot{\rho}(t) = \mathcal{L}\big(\rho(t)\big)$ with linear Liouvillian
$\mathcal{L}$, guiding the system into a steady state $\rho_\infty$ that is diagonal with
respect to the (quasi)energy eigenstates $i=1,2,\ldots, M$ \cite{BreuerPetruccione}. 
The dynamics of the diagonal elements $p_i\equiv\la i|\rho|i\ra$ is given by 
\begin{equation}\label{eq:spr}
\dot{p_i} =  \sum_{j=1}^M ( R_{ij} p_j- R_{ji} p_i),
\end{equation}
with rates $R_{ij}$ for a quantum jump from $j$ to $i$ that, for simplicity, we assume to be
strictly positive, $R_{ij}>0$. 

Now we generalize the single-particle problem (\ref{eq:spr}) 
to a gas of $N$ non-interacting bosons. 
The many-body steady state $\hat{\rho}_\infty$ will be diagonal in the Fock basis $|\bn\ra$
labeled by the occupation numbers $\bn = (n_1,n_2, \ldots n_M)^T$ of the single-particle states 
$i$, obeying $\sum_in_i=N$: $\la\bn'|\hat{\rho}_\infty|\bn\ra = \delta_{\bn'\bn}p_{\bn}$. 
The $N$-boson rate equation
reads
\begin{equation}\label{eq:mpr}
\dot{p}_\bn = \sum_{i,j = 1}^M (R_{ij} p_{\bn_{ji}}-R_{ji} p_\bn)n_i(n_j+1),
\end{equation}
where $\bn_{ji}$ denotes the occupation numbers obtained from $\bn$ by
transferring one particle from $i$ to $j$ \cite{foot:Liouvillian}. 
The steady state with $\dot{p}_\bn =0$ is unique, if every state can be connected to every
other one via a sequence of finite-rate quantum jumps \cite{Hill66,Schnakenberg76}. This is 
true for every $N$, if it is true for the single-particle problem (\ref{eq:spr}).
Equation (\ref{eq:mpr}) is classical in the sense that it involves the diagonal elements of
the density matrix only. However, the bosonic quantum statistics is reflected in the fact
that the rate for a jump from $i$ to $j$ depends both on $n_i$ and $n_j$. This rate reads
$R_{ji}n_i(1+\sigma n_j)$ with $\sigma=-1,0,+1$ for fermions, distinguishable particles, and 
bosons, respectively.

\begin{figure}[t]
\centering
\includegraphics[width=1\linewidth]{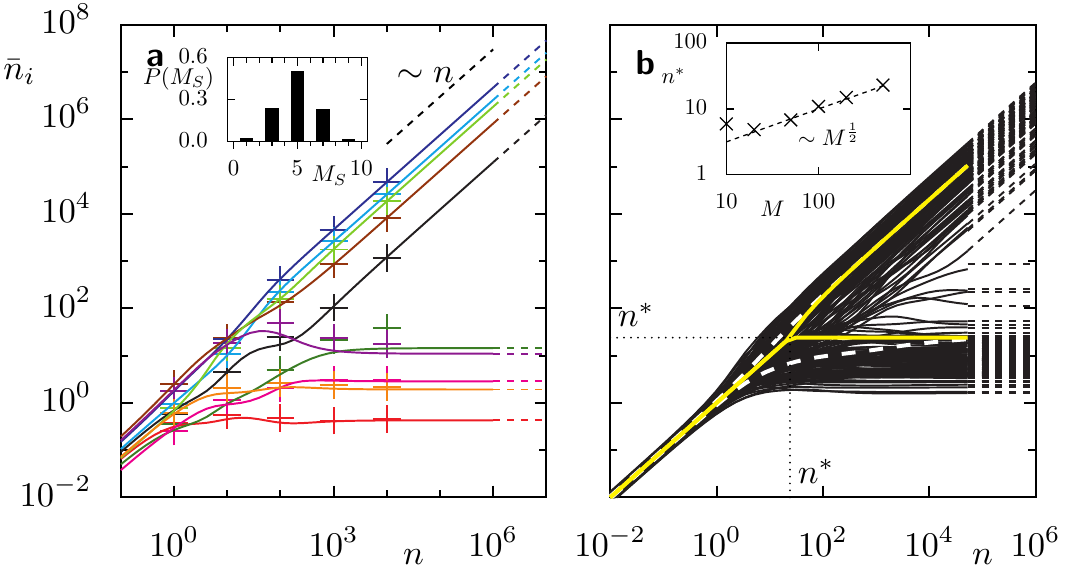}
\caption{\label{fig:bs}
(Color online) 
(a) Mean occupations $\bar{n}_i$ versus density $n=N/M$ for one realization of the
random-rate matrix $R_{ij}$ with $M=10$. Crosses, solid and dashed lines from Monte Carlo, 
mean-field, and asymptotic theory, respectively. For large $n$ the bosonic quantum statistics
leads to the Bose selection of $M_S=5$ states whose occupation grows linearly with $n$. 
Inset: distribution of  $M_S$ for the ensemble of rate matrices.
(b) Like (a), but with $M=200$ and $M_S\approx M/2$. 
Thick lines: average occupation of a selected and a non-selected state, exactly (dashed) and
assuming equal occupation $n$ for $n<n^*$ followed by saturation of the non-selected
occupations (solid). Inset: Crossover density $n^*$ versus system size $M$.}
\end{figure}

As a transparent model system, we will first consider rate matrices $R_{ij}$ given by 
exponentially distributed, independent random numbers \cite{foot:MeanRate}. This choice is 
motivated by the distribution of rates obtained for fully chaotic systems \cite{Supplemental}. 
The rates determine the system completely (and there is no need to define, e.g., the energy of a 
state $i$.)  
In this model the number of states $M$ corresponds to the system size and, thus, the filling factor
$n\equiv N/M$ to the density. 
In Fig.~\ref{fig:bs}, we plot the mean steady-state occupations $\bar{n}_i$ versus 
$n$, for two realizations of $R_{ij}$ with $M=10$ and $M=200$; here $\bar{n}_i=\la\no_i\ra$ 
with number operator $\no_i$ and $\la\cdot\ra=\tr\{\hat{\rho}_\infty\cdot\}$. Crosses result 
from quasi exact Monte-Carlo (MC) simulations \cite{Supplemental,PlenioKnight98}, solid and dashed
lines from the mean-field (MF) and asymptotic theory derived below, respectively. 
The filling $n$ directly controls quantum degeneracy: In the non-degenerate regime $n\ll 1$ 
the relative occupations $\bar{n}_i/N$ approach the $n$-independent single-particle
probabilities $p_i$. Quantum-statistical corrections make themselves felt when entering the
degenerate regime at $n\sim1$. For even larger densities $n$, around a crossover value
$n^*$, we observe that for a group of $M_S$ single-particle states the occupation grows 
linearly with $N$, while the occupation of the remaining states saturates. This is the 
aforementioned effect of Bose selection. Asymptotically, in the ultra-degenerate regime
$n\gg n^*$, the relative occupations of the selected states, $\bar{n}_i/N$, as well as the 
absolute occupations $\bar{n}_i$ of the non-selected states become again independent of $n$. 

Within the ensemble of rate matrices the number of selected states $M_S$ is found to be
always odd (e.g.\ Fig.~\ref{fig:bs}a, inset) and on average $M_S=M/2$ with fluctuations
$\sim M^{1/2}$ (a system with non-extensive $M_S\sim 1$ is presented below).
The crossover density $n^*$ at which the crossover to Bose selection occurs, is 
given by the average number of particles that asymptotically (for large $n$) occupies a
non-selected state. This can be seen by assuming that for $n<n^*$ both groups of states are equally
populated with $n$ particles on average, until at $n=n^*$ the average occupation of the
non-selected states reaches and keeps the saturation value $n^*$ 
(Fig.~\ref{fig:bs}b, thick solid lines). In the random-rate model $n^*$ increases like
$\sim M^{1/2}$ with $M$ (Fig.~\ref{fig:bs}b, inset) \cite{foot:Scaling}. Therefore, in this 
model Bose selection does not occur in the thermodynamic limit, $M\to\infty$ keeping $n$ 
constant, but in finite systems. This behaviour resembles finite-temperature equilibrium Bose
condensation in one dimension.

\begin{figure}[t]
\centering
\includegraphics[width=1\linewidth]{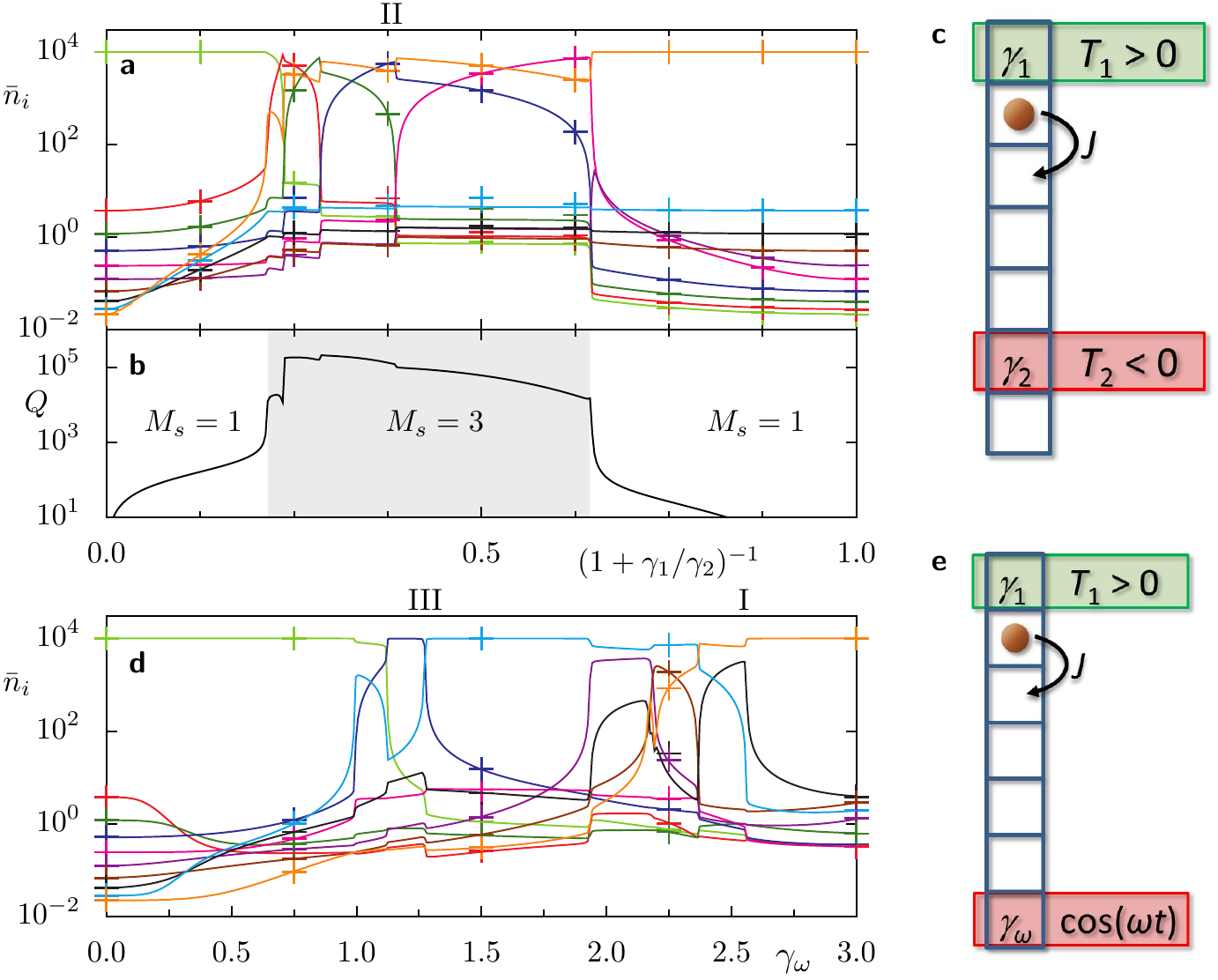}
\caption{\label{fig:chain} 
(Color online) (a) Occupations $\bar{n}_i$ from mean-field (lines) and Monte-Carlo (crosses)
calculations for $N=10^4$ bosons on a tight-binding lattice of $M=10$ sites, weakly coupled
with strengths $\gamma_{1,2}$ to two baths of temperature $T_1=-T_2=J$, as depicted in (c).
(b) Heat flow $Q$ from bath 2 to bath 1 (arbitrary units, $\gamma_1+\gamma_2$ kept constant),
the shaded (unshaded) area corresponds to $M_S=3$ ($M_S=1$). 
(d) Occupations of single-particle Floquet states for the tight-binding chain with
the coupling to bath 2 replaced by a driving term of strength $\gamma_\omega$ as depicted
in (e).}
\end{figure}

In order to treat large systems and to understand the behaviour visible in Fig.~\ref{fig:bs},
we derive a MF theory from the equation of motion $\dot{\bar{n}}_i=\tr(\dot{\hat{\rho}}\no_i)$
for the $\bar{n}_i$ by approximating two-state correlations $\la\no_i\no_j\ra$ by the trivial
ones given by Wick decomposition, $\la\no_i\no_j\ra\approx \bar{n}_i\bar{n}_j$ (for $i\ne j$).
This gives a closed set of non-linear equations for the $\bar{n}_i$,
\begin{equation}\label{eq:mf}
\dot{\bar{n}}_i = \sum_{j=1}^M [R_{ij} \bar{n}_j(\bar{n}_i+1)-R_{ji} \bar{n}_i (\bar{n}_j+1)], 
\end{equation}
and is equivalent to a factorized Gaussian ansatz $\hat{\rho}\propto\exp(-\sum_i\nu_i\no_i)$
with $\nu_i=\ln(\bar{n}_i^{-1}+1)$. The MF theory is confirmed by the MC data
(Figs.~\ref{fig:bs}a, \ref{fig:chain}a, and \ref{fig:chain}d) \cite{foot:Corrections}.

An asymptotic theory for the ultra-degenerate regime, particle number to infinity at
fixed system size, (not to be confused with the thermodynamic limit: system size to infinity at
fixed density) can be derived from the MF Eq.~(\ref{eq:mf}) for $\dot{\bar{n}}_i=0$. The naive
approximation $(\bar{n}_k+1)\simeq \bar{n}_k$ leads to the set of 
equations $0=\bar{n}_i\sum_{j} (R_{ij}-R_{ji})\bar{n}_j$ that generally does not possess a
physical solution with non-negative occupations $\bar{n}_i\ge0$, unless several of the
$\bar{n}_i$ vanish. This gives already a hint why Bose selection occurs, but it does not tell 
us which states become selected, since, e.g., $\bar{n}_i=N\delta_{ik}$ would be a solution for
\emph{any} state $k$. A systematic theory is obtained by assuming that there is some
(yet to be determined) set $S$ of Bose selected single-particle states with occupation
numbers $\sim n$ that are large compared to one as well as to the occupations of the
non-selected states $\sim n^0$. This allows us to expand the $\bar{n}_i$ in powers of 
$n^{-1}$. In leading order we obtain the closed set of 
linear equations for the Bose selected states, 
\begin{equation}\label{eq:selected}
0 = \sum_{j\in S} (R_{ij}-R_{ji})\bar{n}_j, 
\quad 
i\in S.
\end{equation}
The fact that $(R_{ij}-R_{ji})$ is a skew-symmetric matrix guarantees a zero determinant and a
solution of Eq.~(\ref{eq:selected}) provided the set $S$ contains an odd number $M_S$ of states
(for even $M_S$ the existence of a solution requires fine tuned rates $R_{ij}$). 
Thus generically one expects an odd number of selected states, in
accordance with the numerically obtained distribution (Fig.~\ref{fig:bs}a, inset). The next order
describes the occupations of the non-selected states, 
\begin{equation}\label{eq:nonselected}
 \bar{n}_i = \frac{1}{g_i -1} 
\quad\text{with}\quad 
g_i = \frac{\sum_{j\in S}R_{ji}\bar{n}_j}{\sum_{j\in S}R_{ij}\bar{n}_j}, 
\quad i\notin S ,
\end{equation}
and gives also corrections to the occupations of the selected states that we
omit here (even higher orders become essential when allowing also for zero rates, 
$R_{ij}\ge0$ \cite{Supplemental}). 
Equation (\ref{eq:selected}) determines the relative occupations 
among the selected states. These are independent of the total particle number $N$ and, in 
turn, dictate the absolute occupations of the non-selected states via
Eq.~(\ref{eq:nonselected}). The latter, thus, do not depend on the total particle number $N$, 
corresponding to the saturation behaviour visible in Fig.~\ref{fig:bs}.
The total number of particles occupying the selected states, including corrections to the 
leading order (\ref{eq:selected}), is given by $N-\sum_{i\notin S}\bar{n}_i$ 
and increases linearly with $N$ (since the ``depletion'' $\sum_{i\notin S}\bar{n}_i$ is 
independent of $N$). This behavior is generic for the ultra-degenerate regime and generalizes
Bose condensation, where the occupation of a single state $k$ increases with $N$. 
Remarkably, Bose selection is a very consequence of the bosonic quantum statistics and does not
require any assumptions based on equilibrium statistical mechanics.

The set $S$ of selected states has to be determined by the physical requirement that the 
occupations $\bar{n}_i$ of both the selected and the non-selected states are non-negative. 
It can be shown that a unique physical solution with positive occupations exists
\cite{Schomerus13}, as expected from the fact that a unique steady state
of Eq.~(\ref{eq:mpr}) exists (see supplemental material \cite{Supplemental}). 
We are not aware of an easy strategy (beyond trial and error) that
generally allows to determine which and how many states are selected. 
However, if there is a ground-state-like single-particle state $k$, characterized by $R_{ki}>R_{ik}$ for
all $i\ne k$, then only this state $k$ will be
selected and $M_S=1$, corresponding to Bose-Einstein condensation. Namely, since Eq.~(\ref{eq:selected}) is
fulfilled trivially and Eq.~(\ref{eq:nonselected}) gives positive  occupations 
$\bar{n}_{i\ne k}=[R_{ki}/R_{ik}-1]^{-1}>0$ for the non-selected states, this is the (unique) 
physical solution. In contrast, as soon as there is no such ground-state-like state $k$ 
anymore, then more than a single state must be selected. 

An important special case are rate matrices for a system with single-particle energies
$E_1<E_2\le E_3 \cdots$ in weak contact with a thermal bath of inverse temperature $\beta$
for which the rate matrices obey $R_{ji}/R_{ij}=\exp[\beta (E_i-E_j)]$. 
Such rates guarantee detailed balance, i.e.\ the existence of an equilibrium steady state for 
which each summand on the right-hand side of Eqs.~(\ref{eq:spr}) and (\ref{eq:mpr}) 
vanishes independently. In the ultra-degenerate regime, one then recovers from
Eq.~(\ref{eq:nonselected}) the familiar expression
$\bar{n}_i=\{\exp[\beta(E_i-E_1)]-1\}^{-1}$ for $i>1$ while $\bar{n}_1=N-\sum_{i>1}\bar{n}_i$. 
Therefore (excluding ground-state degeneracy $E_1=E_2$) a non-equilibrium steady state
breaking detailed balance, as it is found in driven-dissipative systems, is a necessary
condition for observing Bose selection of more than a single state. However, breaking detailed balance
is not sufficient, as can be inferred from the example of a system driven between two baths of different
positive temperature. In this case the rates sum up $R_{ij}=R^{(1)}_{ij}+R^{(2)}_{ij}$ and,
despite the fact that the combined rates do not lead to detailed balance anymore, they still
obey $R_{1i}>R_{i1}$ for all $i\ne1$ such that only the ground state will be selected. 
Below we will discuss concrete systems of two classes for which $M_S>1$ is found
naturally: (i) systems in weak contact with two baths, one with positive temperature and
another, energy-inverted one with negative temperature, and (ii) time-periodically driven systems
in weak contact with a thermal bath.

Let us now investigate the effect of Bose selection in the concrete physical model 
system of a one-dimensional tight-binding lattice. Such a model describes, e.g., an array of
Josephson junctions, ultracold atoms in optical lattices or vibrons in an ion chain
\cite{BruderEtAl05,BermudezEtAl13}. On the single-particle level, the lattice sites
$\ell=1, \dots,M$ are coupled by tunneling, $\la\ell'|H|\ell\ra=-J \delta_{\ell',\ell\pm1}$ 
with $J>0$. The eigenstates $i$ are delocalized, thus a highly coordinated rate matrix
results from coupling a bath to a local operator like $v_\ell=|\ell\ra\la\ell|$. 
The resulting rate matrix can be derived microscopically within the Born-Markov approximation
\cite{BreuerPetruccione} (see supplemental material for details of the Ohmic baths used here
and a plot of the rate matrix \cite{Supplemental}). 
In order to achieve Bose selection with $M_S>1$ we consider two baths, as sketched in 
Fig.~\ref{fig:chain}c: one with positive temperature $T_1=J$ couples with strength
$\gamma_1$ to $v_1$ and another one with negative temperature $T_2=-J$ couples with strength
$\gamma_2$ to $v_{M-1}$ \cite{foot:symmetry}. Here the negative temperature models a bath 
with occupations that increase with energy.  
In~Fig.~\ref{fig:chain}a we plot the mean occupations of the eigenstates of the tight-binding
chain versus the relative coupling strength $(1+\gamma_1/\gamma_2)^{-1}$ for large filling. 
One can observe Bose selection as a clear separation between highly occupied states on the 
one hand and states with occupations $\lesssim1$ on the other.
For $\gamma_2=0$ the system is in equilibrium and Bose condensation, the selection of a single 
state, is found. When the coupling to the inverted bath is switched on, at 
$(1+\gamma_1/\gamma_2)^{-1}\approx 0.2$ three states become selected 
(Fig.~\ref{fig:chain}a, shaded area). While the data of Fig.~\ref{fig:chain}a correspond to
$M=10$, for $\gamma_2/\gamma_1=1$ we have studied also larger systems with up to $M=300$ sites
and always found 3 states selected. 
This suggests, that the model of Fig.~\ref{fig:chain}c is an example where, in contrast to the
random-rate model, the number of selected states remains of order one (while still being larger
than one). This corresponds to a fragmented condensate with a macroscopic occupation of each 
selected state.
 
As a striking signature for the selection of more than a single state, at the transition from 
$M_S=1$ to $M_S=3$ a significant steady-state heat flow $Q$ from bath 2 to bath 1 is 
established abruptly (Fig.~\ref{fig:chain}b). The heat flow from bath $b=1,2$ into the system
reads $Q_b=\sum_{ij} R^{(b)}_{ji}\bar{n}_i(\bar{n}_j+1)(E_j-E_i)$. This explains an
increase by orders of magnitude from $\sim n$ to $\sim n^2$ when the transition from one to three
selected states occurs [since $\bar{n}_i\sim n$ ($\bar{n}_i\sim 1$) for selected (non-selected) states].
Thus, the mechanism of Bose selection might be used to design quantum devices working far from
equilibrium that allow to switch the heat conductivity via the number of selected states.  

Let us now consider time-periodically driven quantum systems (Floquet systems) 
with Hamiltonian $H(t)=H(t+2\pi/\omega)$ \cite{Shirley65,Sambe73,BreuerHolthaus89}. When 
coupled weakly to a thermal bath, these systems can be described within Floquet-Born-Markov 
theory \cite{BluemelEtAl91,KohlerEtAl97,BreuerEtAl00,HoneEtAl09,KetzmerickWustmann09}. One obtains 
Eqs.~(\ref{eq:spr}) and (\ref{eq:mpr}) with $i$ labeling single-particle Floquet states. In 
the supplemental material we show that the rate differences $R_{ij}-R_{ji}$ are independent 
of the bath-temperature \cite{Supplemental}. According to Eqs.~(\ref{eq:selected}) and
(\ref{eq:nonselected}) this implies that the selected states and their relative occupations
are temperature independent, whereas the occupations of the non-selected states (and thus
also the crossover density $n^*$) are temperature dependent.

Replacing the energy-inverted bath coupled to one end of the tight binding chain
by a coherent periodic driving term $\gamma_\omega J\cos(\omega t)v_M$
with $\hbar\omega=1.5 J$ (Fig.~\ref{fig:chain}e), we obtain the occupations of 
the single-particle Floquet states versus the driving strength $\gamma_\omega$ 
(Fig.~\ref{fig:chain}d). In this driven-dissipative system we observe again both Bose 
condensation into a single state -- which one is controlled by the parameters -- and Bose 
selection of $M_S=3$ states. 

Two more examples that emphasize that Bose selection is a generic and robust effect in open 
time-periodically driven systems are given in the supplemental material \cite{Supplemental}:
the $N$-boson generalizations of the open kicked rotor and the open driven quartic 
oscillator of Ref.~\cite{KetzmerickWustmann10}.

In Figs.~\ref{fig:chain}a and \ref{fig:chain}d, we can study the evolution of the occupations 
with respect to a parameter controlling the rate matrix. Within the asymptotic theory
(\ref{eq:selected}) and (\ref{eq:nonselected}) transitions of states between the groups of
selected and non-selected states are triggered either by the occupation of a selected state
approaching zero or by the
occupation of a non-selected state diverging. Both requires the fine-tuning of a single 
parameter. While at the transition point an even number of states is selected, after the
transition again the generic situation with an odd number of states has to be recovered. 
Thus, a second state has to make a transition at the transition point, too. When approaching
the transition point from the other side, this second state plays the role of the triggering
one. One finds three types of two-state processes, examples of which are labeled by I, II,
and III in Fig.~\ref{fig:chain}a and \ref{fig:chain}d: The transition is triggered from one
side by a selected and from the other one by a non-selected state (I, $M_S$ changes by 2), or
the transition is triggered on both sides either by selected (II) or non-selected (III) states
($M_S$ does not change).

In future work, it will be interesting to study the impact of, e.g., dimensionality, 
particle reservoirs, disorder, and interactions on the effect of Bose selection in
non-equilibrium steady states. A concrete application of Bose selection in a physical system
is the quantum switch for heat conductivity proposed here. 

\begin{acknowledgments}
We warmly thank Martin Holthaus for valuable discussions initiating this work and
Henning Schomerus for providing a proof for the existence of a unique solution of
the asymptotic theory. We acknowledge support through DFG Forschergruppe 760
``Scattering Systems with Complex Dynamics''. 
\end{acknowledgments}


\begin{thebibliography}{35}
\bibitem{foot:Liouvillian}
The many-body Liouvillian generating the dynamics of $\hat{\rho}$ is obtained
from the single-particle one by replacing operators $|i\ra\la j|$ by Fock-space
operators $\hat{b}^\dag_i\hat{b}_j$ with $\hat{b}^\dag_i$ creating a boson in state $i$. 
While the diagonal terms of $\la\bn'|\hat{\rho}|\bn\ra$ solve Eq.~(\ref{eq:mpr}), 
the off-diagonal terms decay.
\bibitem{foot:MeanRate}
The mean rate, here the only free parameter, determines the time scale of the evolution, but
not the steady state.
\bibitem{foot:Scaling}
The asymptotic occupation of a non-selected state (\ref{eq:nonselected}) scales like 
$\bar{n}_i=\sum_{j\in S}R_{ij}\bar{n}_j / \sum_{j\in S} (R_{ji}-R_{ij})\bar{n}_j\approx
\sum_{j\in S}R_{ij} / \sum_{j\in S} (R_{ji}-R_{ij})\sim M^{1/2}$. Here the $\bar{n}_j$ were
first approximated by their average. While in the numerator the sum over rates is then
$\sim M$, the denominator is a sum over rate differences of both signs dominated by
fluctuations, giving $\sim M^{1/2}$ only.
\bibitem{foot:Corrections} 
Residual deviations that are visible for some data points of the non selected states are
largely cured when MF theory is augmented by keeping also non-trivial two-state correlations
like $\la(\no_i-\bar{n}_i)(\no_j-\bar{n}_j)\ra$ for $i\ne j$, while neglecting non-trivial
three-state correlations $\la(\no_i-\bar{n}_i)(\no_j-\bar{n}_j)(\no_k-\bar{n}_k)\ra$. The 
mechanism of Bose selection is not altered by such corrections.
\bibitem{foot:symmetry} 
The symmetric choice $v_{M}$ is special and always leads to $M_S=1$ with either the
state of lowest or largest energy selected.
\end{thebibliography}

\section{Supplemental Material}

\subsection{Quantum-jump Monte-Carlo simulations}
Even though the master equation (2) involves only the Fock states and not the full 
many-body Hilbert space containing also their coherent superpositions, the problem still 
grows exponentially with both particle number $N$ and system size $M$. Therefore, in order to 
compute exact steady-state expectation values from Eq.~(2), we have to resort
to quantum-jump-type Monte-Carlo (MC) simulations (crosses in Figs.~1 and 2) 
[40]: We generate a random process in the classical space of sharp occupation
numbers $\bn$. Namely, according to the sum and the relative weight of the many-body rates
$R_{ij}(n_i+1)n_j$ leading away from the current state $\bn$, we draw both the time after
which a quantum jump happens and the new state $\bn_{ij}$, respectively. 
Expectation values like $\bar{n}_i$ are computed by averaging over a random path. The Monte-Carlo
method gives quasi exact results, in the sense that the accuracy is controlled by the length
of the random path. 


\subsection{Allowing for zero rates}
It is instructive to release the restriction $R_{ij}>0$ to $R_{ij}\ge 0$, still assuming
that every state $i$ can be reached from every other state $j$ via a sequence of finite-rate
quantum jumps. In this case the set of selected states cannot consist of different subsets
with zero mutual rates, as this would leave the relative occupation of the subsets 
undetermined. Moreover, the leading contribution to the asymptotic occupation $\bar{n}_i$ of
a non-selected state $i$ with rates $R_{ij}=0$ for all selected states $j$ appears in a higher
order of our expansion (but is still independent of $N$ such that the occupations of the
non-selected states saturate for large $N$).
This suggests that if there is one class of rates that are non-zero, but very small compared 
to the other rates, then the occupation numbers $\bar{n}_i$ can be dominated by higher-order
corrections, before for very large $n$ they eventually reach their asymptotic values given by
Eqs.~(4) and (5)

\subsection{Existence and uniqueness of asymptotic mean-field solution}
Here we present proofs due to Henning Schomerus, both for the existence and the uniqueness
of a physical solution of Eqs.~(4) and (5) for a generic rate matrix. 

The rate imbalances $R_{ij}-R_{ji}$ form a skew-symmetric matrix $A=R-R^T=-A^T$. Let the vector
$\bm{\nu}=(\nu_1,\nu_2,\ldots,\nu_M)^T$ describe the occupations of the single-particle states in the leading 
order of our expansion in powers of $n^{-1}$. In this order only the selected states $i\in S$ have non-zero 
positive occupation numbers $\nu_i > 0$, whereas $\nu_i=0$ for $i\notin S$. We can define the vector
$\bm{\mu}=A\bm{\nu}$. According to Eq.~(4) one has $\mu_i=0$ for $i\in S$ and, moreover, according to Eq.~(5)
one has to require $\mu_i<0$ for $i\notin S$ in order to have finite, positive occupations for the
non-selected states. 
Thus, the problem of finding an asymptotic mean-field solution can be 
reduced to the problem of finding a set $S$ of selected states such that
$\bm{\mu}=A\bm{\nu}$ with $\nu_i>0$ and $\mu_i=0$ for $i\in S$ and $\nu_i=0$ and $\mu_i<0$ for
$i\notin S$. 
(Note that we exclude the non-generic situation with $\nu_i=\mu_i=0$ for some $i$ that
requires fine-tuned rate matrices. This situation, however, is relevant at transition 
points, where the set of selected states changes in response to a parameter variation.)

\paragraph{Uniqueness of set $S$:} Assume there exist two different sets $S_1$ and $S_2$ both leading to physical 
solutions $\bm{\nu}_1$ and $\bm{\nu}_2$ giving rise to non-negative occupation numbers. 
Then, it follows that $0\ge \bm{\nu}_2^T\bm{\mu}_1 = - \bm{\nu}_1^T\bm{\mu}_2\ge0$, using
$\bm{\nu}_2^T\bm{\mu}_1=\bm{\nu}_2^TA\bm{\nu}_1=(\bm{\nu}_2^TA\bm{\nu}_1)^T=\bm{\nu}_1^TA^T \bm{\nu}_2=
-\bm{\nu}_1^TA\bm{\nu}_2=- \bm{\nu}_1^T\bm{\mu}_2$. 
However, this requires that both $\bm{\nu}_2^T\bm{\mu}_1 =0$ and $\bm{\nu}_1^T\bm{\mu}_2=0$, such that 
$S_2\subset S_1$ \emph{and} $S_1\subset S_2$, respectively, leading to the contradiction $S_1=S_2$. 

\paragraph{Existence of set $S$:} Restrict $S$ to sets comprising an odd number $M_S$ of states (generically even 
$M_S$ do not allow for solutions, as explained in the main text). 
Each choice of $S$ gives rise to a possibly non-physical solution $\tilde{\bm{\nu}}$ with $\tilde{\bm{\mu}}=A
\tilde{\bm{\nu}}$. The vector of signs $\bf{\sigma}$ with $\sigma_i=\text{sign}(\tilde{\nu}_i)$ if $i\in S$ 
and $\sigma_i=-\text{sign}(\tilde{\mu}_i)$ if $i\notin S$, with the overall sign fixed by the convention
$\sigma_1=+1$, distinguishes physical solutions ($\sigma_i=+1$ for all $i$) from non-physical solutions. 
Now one can observe: (i) Each vector $\sigma$ occurs at most once. Namely, if $S_1$ and $S_2$ gave rise to the 
same vector $\bf{\sigma}$ then the modified rate imbalance matrix $\tilde{A}_{ij}=\sigma_iA_{ij}\sigma_j$ had the
two physical solutions $S_1$ and $S_2$ in contradiction to the uniqueness. (ii) The number $2^{M-1}$ of possible 
vectors $\bf{\sigma}$ equals the number $\sum_{M_S=1,3,\ldots} {M\choose M_S}=2^{M_S-1}$ of possible sets $S$. 
Therefore, each vector $\bf{\sigma}$ occurs once, in particular the one with $\sigma_i=+1$ for all $i$ 
leading to the physical solution with positive occupations.

\subsection{Floquet-Born-Markov theory}
A quantum Floquet system, characterized by a time-periodic Hamiltonian $H(t)=H(t+2\pi/\omega)$,
possesses solutions of the form $|\varphi_i(t)\ra=\re^{-i\varepsilon_i t/\hbar}|u_i(t)\ra$, 
with quasienergies $\varepsilon_i$ and time-periodic Floquet modes
$|u_i(t)\ra=|u_i(t+2\pi/\omega)\ra$ forming a complete orthonormal basis at every instant in 
time [46-48]. 
When coupled weakly to a thermal bath, according to Floquet-Markov theory 
[49-52], the steady state is given by a
time-periodic density matrix $\rho(t)=\sum_i p_i |u_i(t)\ra\la u_i(t)|$ 
that is diagonal in the Floquet modes and characterized by the time-independent probabilities 
$p_i$ obeying a master equation of the form (1). We assume a bath 
of harmonic oscillators $\alpha$, with angular frequencies $\omega_\alpha$ and 
annihilation operators $a_\alpha$, that couples to the system operator $v$ like
$v\sum_\alpha c_\alpha (a_\alpha+a_\alpha^\dag)$.  Then the rates read  
\begin{equation*}
R_{ij}
=\frac{2\pi}{\hbar}\sum_{m=-\infty}^\infty|v_{ji}(m)|^2 
g(\varepsilon_j-\varepsilon_i-m\hbar\omega)
, \end{equation*}
with $v_{ji}(m)=\frac{\omega}{2\pi}\int_0^{2\pi/\omega}\!\rd t\, 
\re^{-im\omega t}\la u_j(t)|v|u_i(t)\ra$ and bath correlation function
$g(E)=J(E)[\exp(\beta E)-1]^{-1}$. The latter is determined 
by the inverse temperature $\beta$ and the spectral density
$J(E)=\sum_\alpha c_\alpha^2 [\delta(E-\hbar\omega_\alpha)-\delta(E+\hbar\omega_\alpha)]$ that
we assume to be ohmic, $J(E)\propto E$. 
The Floquet gas of $N$ non-interacting bosons is described by Eq.~(2) with
$\bn$ denoting the occupation numbers of the single-particle Floquet modes.

Remarkably, we obtain the temperature-independent expression 
\begin{equation*} 
R_{ij}-R_{ji}=\frac{2\pi}{\hbar} 
\sum_{m=-\infty}^\infty|v_{ji}(m)|^2J(\varepsilon_j-\varepsilon_i-m\hbar\omega)
\end{equation*} 
for the rate differences. This implies that for a periodically driven system the set of
selected states does not depend on the bath temperature. In particular, according
to Eqs.~(4) and (5) the relative occupations among the selected states do not depend on
temperature, whereas the occupations of the non-selected states (and thus also the crossover
density $n^*$) are temperature dependent. 

In the limit of vanishing periodic forcing (i.e. for a system described by a time-independent 
Hamiltonian) the Floquet modes and quasienergies approach the eigenmodes and eigenenergies of the 
time-independent Hamiltonian, such that $v_{ji}(m)$ is non-zero only for $m=0$. As a
consequence the rates $R_{ij}$ are given by a single term only 
\begin{equation*}
R_{ij}
=\frac{2\pi}{\hbar}|v_{ji}(0)|^2 
g(\varepsilon_j-\varepsilon_i-m\hbar\omega)
, \end{equation*}
and, thanks to the general property $g(-E)=\re^{\beta E}g(E)$ of the bath correlation function,
give rise to detailed balance, with $R_{ij}/R_{ji}=\exp[\beta(E_j-E_i)]$. In  contrast, for
finite periodic driving detailed balance is not guaranteed, because terms with different $m$
contribute to the rates. In a non-driven system detailed balance can be broken by
coupling the system weakly to different baths of different temperature; in this case the rates
resulting from the different baths simply sum up, $R_{ij}=R_{ij}^{(1)}+R_{ij}^{(2)}+\cdots$.

The assumption of a diagonal steady-state density matrix,
$\rho_\infty = \sum_i p_i|i\ra \la i|$ or 
$\hat{\rho}_\infty=\sum_{\bm{n}}p_{\bm{n}}|\bm{n}\ra\la\bm{n}|$,
for weak enough coupling to a heat bath has to be discussed
carefully for Floquet systems.
In these systems all quasienergies can be
placed in a finite energy interval of size $\hbar \omega$,
leading to smaller and smaller near degeneracies 
as the size $M$ of the single-particle state space is increased.
The assumption of a diagonal density matrix is fulfilled
only, if for any two states, their coupling provided by the heat bath
is smaller than their quasienergy splitting.
In the limit $M \to \infty$ this assumption cannot be guaranteed
and the consequences for single particle systems
and their density matrix have been discussed in 
Ref.~[52]. 

For many-particle systems, even when $M$ is fixed, the state space
increases exponentially with the particle number $N$ and therefore 
the size of the smallest quasienergy splitting 
decreases exponentially. Quickly it will be much smaller
than even the tiniest experimentally realizable coupling
to a heat bath. This could destroy the diagonality
of the density matrix for large particle numbers $N$.
This is not the case, however, for the here studied 
non-interacting gas. There all transitions between many-particle
Floquet states can be reduced to single-particle transitions for 
which the single-particle quasienergy splittings are relevant. 
Therefore we can make the assumption of a diagonal density
matrix for a non-interacting gas.

A switching mechanism due to an avoided
quasienergy crossing in a single-particle system was
reported in Ref.~[53]. It allowed to switch the asymptotic steady
state between the lower and the upper well
of an asymmetric double-well potential by a periodic driving,
that is much weaker than the asymmetry. It works even if
the Floquet states involved in the avoided crossing 
have just a small occupation. It would be interesting to study
how such a switching mechanism due to an avoided crossing is
reflected in the corresponding ideal Bose gas and how it affects
the set of Bose selected states.

\begin{figure}[t!]
\centering
\includegraphics[width=1\linewidth]{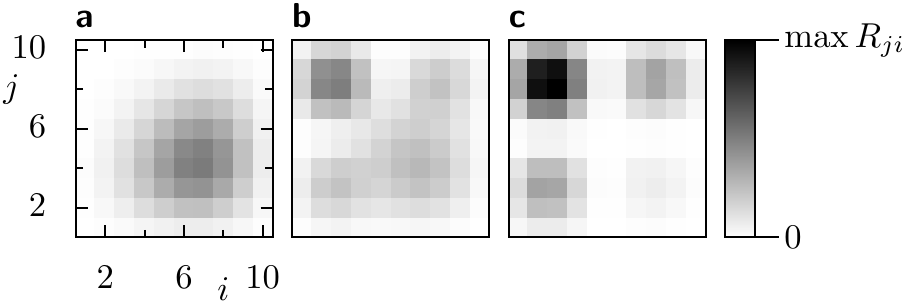}
\caption{\label{fig:rates} 
Rates $R_{ji}$ for a quantum jump from state $i$ to state $j$ for the tight-binding chain in contact with two 
heat baths (Figs.~2a-c in the main text). From left to right the three plots correspond to the system coupled 
to the positive temperature bath only, coupled to both baths with equal strength, coupled to the bath with
negative temperature only, respectively.}
\end{figure}

\begin{figure}[t!]
\centering
\includegraphics[width=1\linewidth]{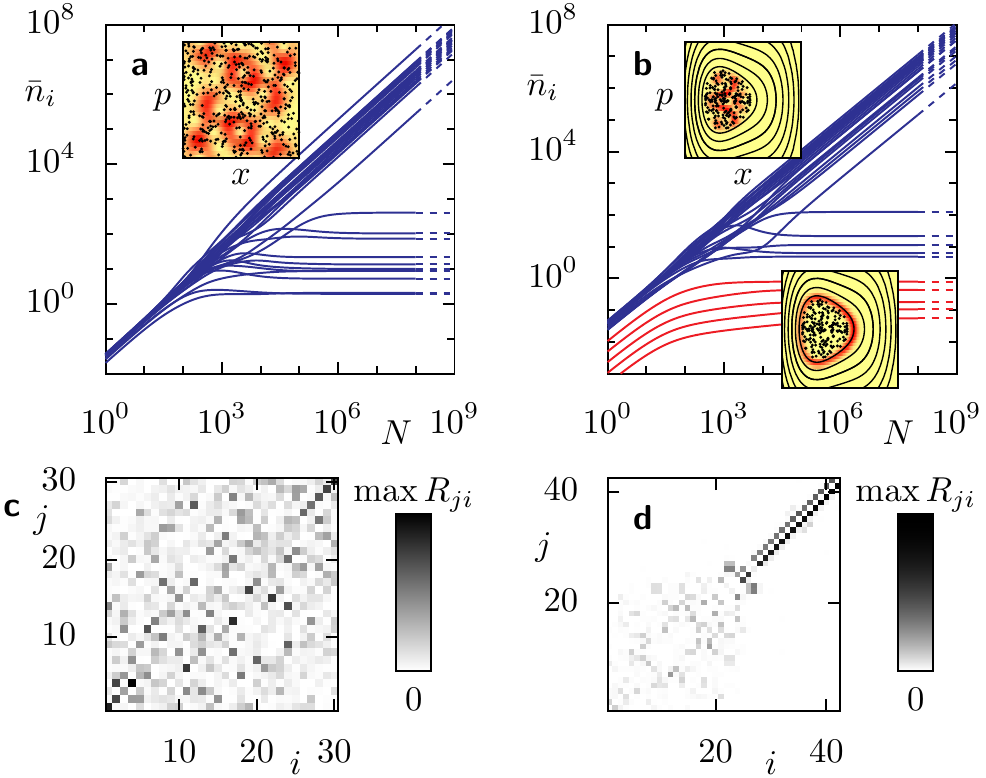}
\caption{\label{fig:floquet} 
Mean occupations $\bar{n}_i$ versus total particle number $N$ for a fully chaotic kicked
rotor ({a}) and a driven quartic oscillator with mixed phase space ({b}) coupled to a thermal
bath. The corresponding rate matrices are given in subfigures (c) and (d), respectively. 
Both (a) and (b) show Bose selection for large $N$. Red (blue) lines indicate regular 
(chaotic) states, solid (dashed) lines result from mean-field (asymptotic) theory. Insets:
Stroboscopic Poincar\'e plots of the classical dynamics in phase space in comparison to the
Husimi distribution of a representative chaotic and regular quantum state (red shaded).}
\end{figure}

\subsection{Rate matrices for the tight-binding chain}
In Fig.~S\ref{fig:rates} we plot the rate matrix used for the tight-binding chain coupled to two heath baths of 
positive and negative temparature, as it is described in Figs.~2a-c of the 
main text. From left to right the three plots correspond to the system coupled only to the positive temperature 
bath, coupled to both baths with equal strength, coupled only to the bath with negative temperature, 
respectively. For the leftmost (rightmost) case simply the state $k$ with the lowest (largest) energy is 
selected, since for this state $R_{ki}>R_{ik}$. For the middle case with coupling to both baths we find three 
selected states, namely the states 4, 5 and 10 (labeling the states by $i=1,2,\ldots,M$ from the lowest to the 
highest energy). The selected states 4 and 5 are characterized by low rates to states of higher energy and from 
states with lower energy. However, we are not aware of a simple general strategy for guessing the selected sates simply from looking at the rate matrix.

\subsection{Two further examples for Bose selection in time-periodically driven model systems}
In order to emphasize the fact that Bose selection is a generic and robust effect in 
driven-dissipative ideal Bose gases, let us study the phenomenon also in the two 
time-periodically driven model systems in weak contact with a thermal bath that,
on the single-particle level, were discussed in detail in Ref.~[54]. 
The first one is the kicked rotor with parameters giving $M=30$ Floquet modes that in the 
semiclassical sense are all chaotic. The resulting rates $R_{ij}$ (plotted in Fig.~S\ref{fig:floquet}c) 
appear to be exponentially distributed and give rise to the Bose selection of $M_S=19$ states, as shown in
Fig.~S\ref{fig:floquet}a. The second system (Fig.~S\ref{fig:floquet}b) is a periodically driven 
quartic oscillator having a mixed phase space with a central region of 26 chaotic states, 
surrounded by infinitely many regular states that can be labelled by $i=27,28, \ldots$ in 
radial direction. Bose selection occurs among the chaotic states that are connected by
random-like highly coordinated rates (the rate matrix is plotted in Fig.~S\ref{fig:floquet}d); 
21 states are 
selected. In turn, the regular states are coupled essentially to their nearest neighbor only and
their occupation decays rapidly with increasing $i$, allowing us to truncate modes with $i\ge 32$.

\end{document}